\documentclass[a4paper]{PoS}

\usepackage[utf8]{inputenc}
\usepackage{amsmath,amssymb,bm}
\usepackage[numbers,sort&compress]{natbib}
\usepackage[usenames,dvipsnames]{xcolor}
\usepackage[hidelinks,colorlinks,allcolors=Blue]{hyperref}

\newcommand{\pysecdec}{\texttt{py{\textsc{SecDec}}}}
\newcommand{\secdec}{\texttt{{\textsc{SecDec}}}}

\title{Numerical evaluation of multi-loop integrals}

\ShortTitle{Numerical evaluation of multi-loop integrals}

\author{\speaker{Stephan Jahn}\\
        Max-Planck-Institut für Physik, München, Germany\\
        E-mail: \email{sjahn@mpp.mpg.de}}

\abstract{
We present updates on the development of \pysecdec{}, a toolbox
to numerically evaluate parameter integrals in the context of
dimensional regularization. 
We discuss difficulties with loop integrals in the special
kinematic condition where the squared momentum of a leg is equal to the squared mass of
a propagator. 
We further discuss some features of a Quasi Monte Carlo (QMC) integrator
that can optionally run on Graphics Processing Units (GPUs).
}

\FullConference{Loops and Legs in Quantum Field Theory (LL2018)\\
		29 April 2018 - 04 May 2018\\
		St. Goar, Germany}

\begin{document}

\section{Introduction}

Current and future searches for New Physics at the LHC are focused on precision measurements. Higher order calculations are crucial for the interpretation of the data.
A well established procedure for multi-loop calculations is: (i) draw all contributing Feynman diagrams, (ii) insert the Feynman rules to produce
algebraic expressions, (iii) identify the loop integrals to be computed, (iv) reduce these integrals to a set of master integrals, and (v) solve the
master integrals.

Sector decomposition~\cite{Binoth:2000ps,Heinrich:2008si} provides an automated way to evaluate the master integrals numerically.
In this article, we present recent developments in \pysecdec{}~\cite{Borowka:2017idc}, the successor of the program \secdec{}~\cite{Carter:2010hi,Borowka:2012yc,Borowka:2015mxa}.
Other public sector decomposition tools are \texttt{sector\_decomposition}~\cite{Bogner:2007cr} supplemented with \texttt{CSectors}~\cite{Gluza:2010rn}
and \texttt{FIESTA}~\cite{Smirnov:2008py,Smirnov:2009pb,Smirnov:2013eza,Smirnov:2015mct}.

Given the amplitude in terms of master integrals, the major bottleneck after factorization of the poles in the dimensional regulator is the numerical integration.
Suitable numerical integration algorithms are crucial to achieve reliable results in an acceptable time frame. The benefits of Quasi Monte Carlo (QMC) integration (see e.g.~\cite{dick_kuo_sloan_2013})
in combination with sector decomposition and running the numerical integration on Graphics Processing Units (GPUs) have first been pointed out in~\cite{Li:2015foa}.
The method has also been successfully applied to phenomenological calculations, e.g. Higgs-boson pair production~\cite{Borowka:2016ehy,Borowka:2016ypz} and
H+jet production~\cite{Jones:2018hbb} at NLO. Considering these successes as a proof of principle, we announce an upcoming version of \pysecdec{} with a
builtin Quasi Monte Carlo integrator that can optionally run on nvidia GPUs using CUDA.

We further address problems that can occur when integrals in special kinematic configurations are considered.
There are integrals that could not be dealt with by either of the public sector decomposition programs~\cite{Dubovyk:2016zok}.
The simplest possible example that triggers the problem is a regulated $x-y$ in the denominator of the integrand, where $x$ and $y$ are integration variables.
We therefore refer to it as the ``$x-y$'' problem.

The structure of this article is as follows: The upcoming Quasi Monte Carlo integrator is described in section~\ref{sec:qmc}.
We describe the previously introduced ``$x - y$'' problem and its remedy in \pysecdec{} in section~\ref{sec:xminusy}
before we conclude in section~\ref{sec:conc}.

\section{Quasi Monte Carlo}
\label{sec:qmc}

Consider the $d$ dimensional integral $\mathrm{I}_d$,
\begin{equation}
\mathrm{I}_d [ f ] \equiv \int_0^1 \, \mathrm{d}x_1 ... \int_0^1 \, \mathrm{d}x_d \  f(\mathbf{x}), \  \mathbf{x} = \left(x_1,...,x_d\right)
\label{eq:qmc_integral}
\end{equation}
of a function $f:\  \left[0,1\right]^d \rightarrow \mathbb{K},\  \mathbb{K} \in \{\mathbb{R},\mathbb{C}\}$ over the unit hypercube.
Following~\cite{Li:2015foa}, we use shifted rank-1 lattice rules to obtain an unbiased estimate of $\mathrm{I}_d$ from $n$ fixed sampling points
$\lbrace{}\mathbf{x}_j: \mathbf{x}_j \in \left[ 0,1 \right] ^ d, \  j=1,...,n\rbrace$
and $m$ random shifts $\lbrace{}\boldsymbol{\Delta}_k: \boldsymbol{\Delta}_k \in [ 0,1 ) ^ d, \  k=1,...,n\rbrace$ as
\begin{align}
&\mathrm{I}_d[f] \approx \bar{Q}_{d,n,m}[f] \equiv  \frac{1}{m} \sum_{k=1}^{m} Q_{s,n}^{(k)}[f], &
&Q_{d,n}^{(k)}[f] \equiv \frac{1}{n} \sum_{j=1}^{n} f \left( \left\{ \frac{(j-1) \mathbf{g}}{n} + \boldsymbol{\Delta}_k \right\} \right),&
\label{eq:integral_approx_general}
\end{align}
where $\mathbf{g} \in \mathbb{N}^d$ is the so-called generating vector and $\lbrace \rbrace$ means taking the fractional part. An unbiased error estimate is given by
\begin{equation}
\sigma^2_{d,n,m}[f] \equiv \mathrm{Var} [ \bar{Q}_{d,n,m} [f] ] \approx \frac{1}{m(m-1)} \sum_{k=1}^{m} ( Q_{s,n}^{(k)}[f] - \bar{Q}_{s,n,m}[f] )^2.
\label{eq:qmc_error_estimate}
\end{equation}
An extensive review of the QMC method including proofs of the theorems we use in the following paragraphs can be found e.g. in~\cite{dick_kuo_sloan_2013}.

It is proven that the error of the QMC estimate of the integral asymptotically scales with the number of lattice points $n$ as $\mathcal{O}(1/n^{s+1})$, where $s$ is the smoothness
of the periodically continued integrand\footnote{In fact, the sth derivative of $\tilde{f}$ must be \emph{absolutely continuous}, not only \emph{continuous} as usually suggested by the notation $C^{s}$.}
\begin{align}
&\tilde{f} \in C^s(\mathbb{R}^d \rightarrow \mathbb{K}), &
&\tilde{f}(\mathbf{x}) \equiv f(\lbrace\mathbf{x}\rbrace).&
\end{align}
Sector decomposed functions are typically infinitely smooth (any derivative is well-defined and finite) but not periodic. However, there are a number of periodizing transforms
implemented via a substitution of the integration variables $\mathbf{x} \rightarrow \mathbf{z}(\mathbf{x})$.
In these proceedings, we consider the Korobov transform of weight $\alpha$,
\begin{align}
&z_j(x_j) = \int_0^{x_j} \mathrm{d} u \ \omega(u),&
&\omega(u)= \frac{u^\alpha(1-u)^\alpha}{\int_0^1 \mathrm{d} u' \  {u'}^\alpha (1-u')^\alpha},&
\label{eq:korobov}
\end{align}
and the baker's transform,
\begin{equation}
z_j(x_j) = 1-\left| 2 x_j - 1 \right|.
\label{eq:baker}
\end{equation}

The Korobov transform forces the integrand to go to zero everywhere on the boundary.
Assuming a $C^\infty$ nonperiodic integrand $f$ and an integer weight $\alpha \in \mathbb{N}$, the Korobov transformed
and periodically continued integrand $\tilde{f}^{K_\alpha}$ is $C^{\alpha-1}$ for odd $\alpha$ and $C^{\alpha}$ for even $\alpha$. This can be seen by examining the derivative
of $\tilde{f}^{K_\alpha}$ on the border of the unperiodized transformed integrand $f^{K_\alpha}$: One finds that the periodization of a high-enough derivative becomes discontinuous.
Hence, we expect the relative error of sector decomposed integrals to asymptotically scale like $\mathcal{O}(1/n^\alpha)$ (odd $\alpha$) or $\mathcal{O}(1/n^{\alpha+1})$
(even $\alpha$) with the number of QMC lattice points $n$.

The baker's transform periodizes the integrand by mirroring rather than forcing it to a particular value on the integration boundary.
Note that the baker's transform is only $C^0$ which naively suggests an asymptotic $\mathcal{O}(1/n)$ scaling. However,
the asymptotic scaling of the error can be proven to be $\mathcal{O}(1/n^2)$ by considering the baker's transform as a modification of the lattice
rather than an integral transform. Further note that the derivative of the baker's transform (where it exists) is plus or minus 2, which leads
to a (piecewise) constant Jacobian factor when considered as an integral transform.

\begin{figure}[htbp]
\hfill \\
\begin{center}
\includegraphics[width=0.3\textwidth]{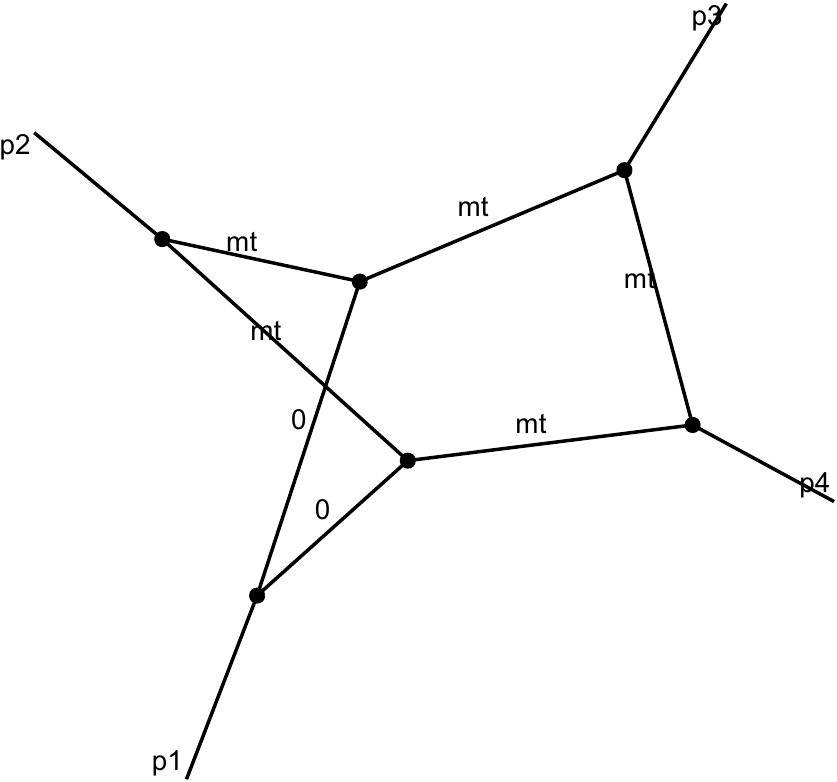}
\end{center}
\caption{A massive nonplanar double box that appears as a master integral in the gluon channel of two Higgs boson production~\cite{Borowka:2016ehy,Borowka:2016ypz}.}
\label{fig:HHNP2b}
\end{figure}

\begin{figure}[htbp]
\hfill \\
\begin{center}
\includegraphics[width=\textwidth]{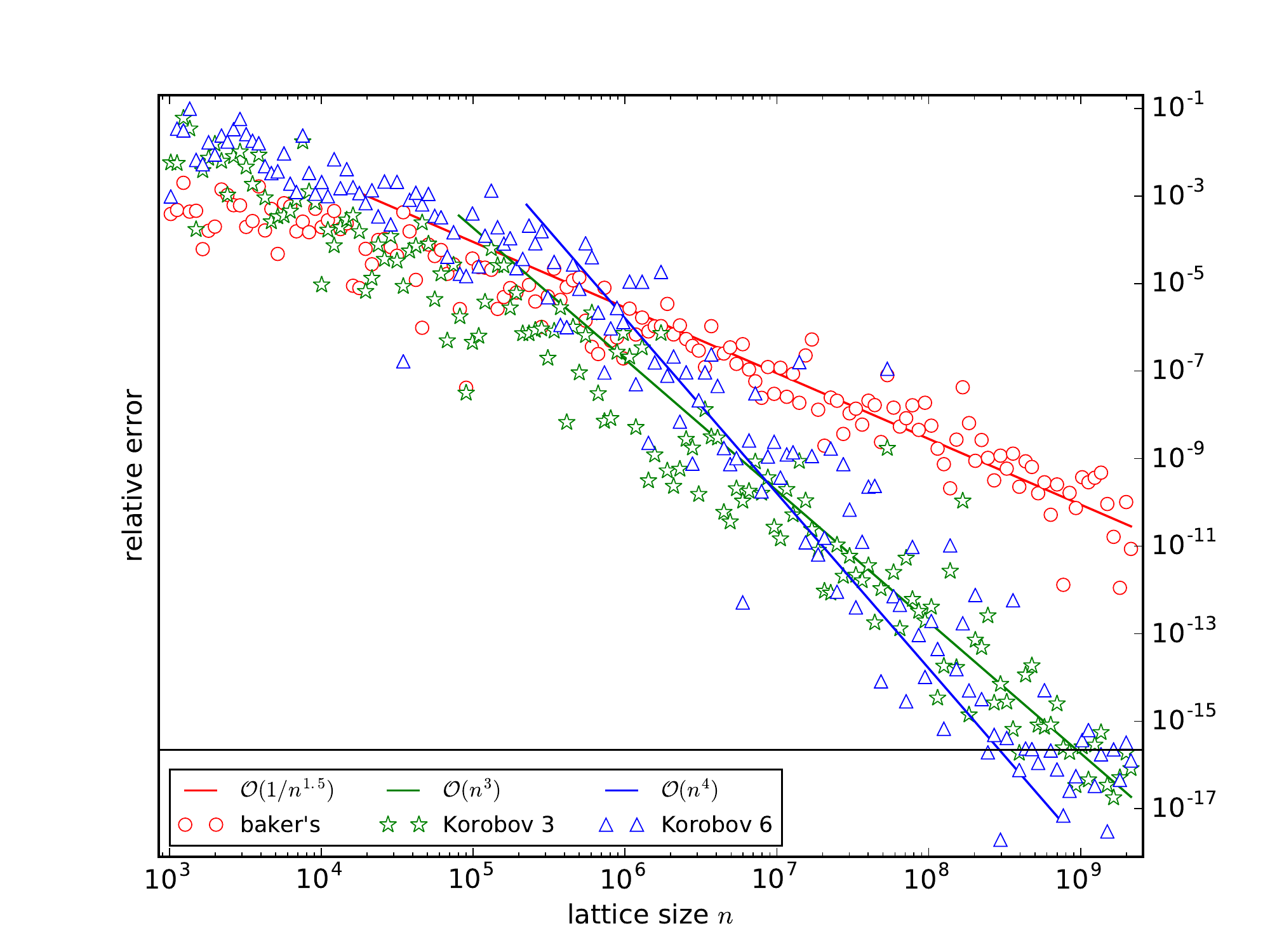}
\end{center}
\caption{Scaling behavior of the real finite part of the double box integral depicted in Figure~\ref{fig:HHNP2b}
integrated with the QMC method using different periodizing transforms. The red circles show
the reported relative errors versus the number of points $n$ in the QMC lattice using the baker's periodizing transform.
The observed scaling of the error roughly matches $\mathcal{O}(1/n^{1.5})$. The green stars show the reported relative errors of the same integral
using a Korobov transform with weight 3. Due to the higher smoothness, a better scaling of $\mathcal{O}(1/n^3)$ is achieved.
The blue triangles show the reported relative error of the integral using a Korobov transform with weight 6.
The scaling matches about $\mathcal{O}(1/n)$ below $n \lesssim 10^6$ and improves to $\mathcal{O}(1/n^4)$ for $n \gtrsim 10^6$. The black
line marks the precision of the floating point type (double) used in the setup to generate this plot.}
\label{fig:scaling_HHNPb2}
\end{figure}

As an example integral, we consider the 2loop nonplanar box depicted in Figure~\ref{fig:HHNP2b}
which appears as a master integral in two Higgs boson production~\cite{Borowka:2016ehy,Borowka:2016ypz}.
The number of lattice points $n$ against the relative error of the real finite part is shown in Figure~\ref{fig:scaling_HHNPb2}.

It turns out that the overall scaling of the error is better than $\mathcal{O}(1/n)$ in all cases. However, the scaling is very noisy in the sense that
lattices of similar size can lead to error estimates that differ by orders of magnitude. We checked this to be an intrinsic property of the lattice applied
to a particular function, i.e. it is \emph{not} due to statistical fluctuations when estimating the error. We further observe that the overall scaling may improve
with increasing lattice size $n$. For example, if the double-box integral depicted in Figure~\ref{fig:HHNP2b} is computed with a Korobov transform of weight 6 then
the scaling improves at $n\approx10^6$, see Figure~\ref{fig:HHNP2b}.

Due to the noise, it is hard to measure the scaling. Fitting a single line over the entire range
does not make sense because the scaling clearly changes at certain values of $n$. Fitting lines to only part of the
points would lead to somewhat arbitrary results depending on the considered ranges. We therefore only plot lines with predefined
scaling behavior for guidance - they are \emph{not} to be understood as fits.
It is therefore hard to judge if we see the Korobov transform of weight 6 scale as the theoretically expected $\mathcal{O}(1/n^7)$
in the tail of the distribution or not. We expect to observe a $\mathcal{O}(1/n^7)$ scaling with larger lattices which are out of reach for
practical reasons such as the machine precision and the available compute power.

\begin{figure}[htbp]
\hfill \\
\begin{center}
\includegraphics[width=\textwidth]{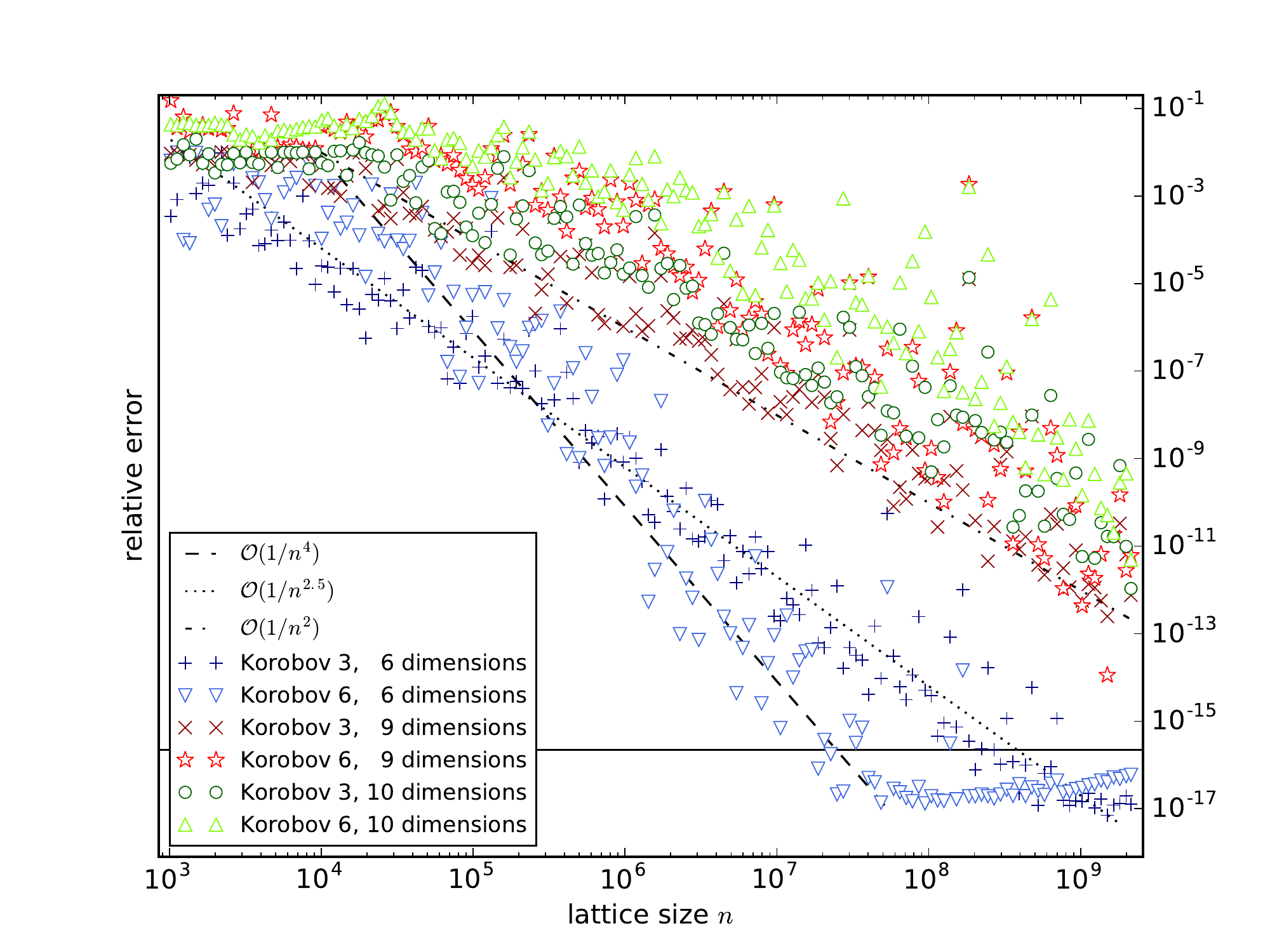}
\end{center}
\caption{Scaling behavior of the Korobov transforms with weight 3 and 6 in different dimensions.
The dashed, dotted, and dash-dotted lines show $\mathcal{O}(1/n^{4})$, $\mathcal{O}(1/n^{2.5})$, and $\mathcal{O}(1/n^{2})$
scaling normalized to arbitrary starting points.
The blue pluses (triangles) show the reported error when integrating a constant using a Korobov with weight 3 (6) against the number
of QMC lattice points $n$ in 6 dimensions.
The Korobov transform of weight 3 roughly shows a $\mathcal{O}(1/n^{2.5})$ to $\mathcal{O}(1/n^{3})$ scaling starting from the smallest
lattices $n\approx{}10^3$ used. The Korobov transform of weight 6 exhibits a better $\mathcal{O}(1/n^4)$ scaling for $n\gtrsim{}10^4$ but
yields an error that is roughly independent of the lattice size for smaller lattices. A similar behavior is observed
for both Korobov transforms, with weights 3 and 6, in 9 and 10 dimensions.}
\label{fig:scaling_one}
\end{figure}

We further observe the scaling to worsen for higher dimensional integrals. Part of the problem seems to be the periodizing transforms~\cite{Kuo2007}.
As a test, we integrate the constant $1$ using the Korobov transform with different weights in different dimensions. The results are shown in Figure~\ref{fig:scaling_one}.
The baker's transform is not shown in that plot, because it would have zero error for any lattice since it does not add variance to the original (constant) integrand.
As an interesting feature, we find that Korobov transforms of higher weight or higher dimensions show a plateau where the error is approximately independent of the
lattice size $n$ for small $n$. However, the error starts to decrease with the lattice size for large enough lattices. We further observe lower errors for lower
weight Korobovs and relatively small lattices. This suggests that ``high'' dimensional integrals may in practice be integrated to higher accuracy with worse scaling
transforms due to available compute power.

\section{The ``$x - y$'' problem}
\label{sec:xminusy}

Consider an integral of the form
\begin{equation}
\int_0^1 \, \mathrm{d}x_1 ... \int_0^1 \, \mathrm{d}x_d \  \prod_{j=1}^m P_j \left( \lbrace x \rbrace, \lbrace u \rbrace \right) ^ {a_j + b_j \varepsilon},
\label{eq:general_integrand}
\end{equation}
where the $P_j$ are polynomials of the integration variables $x_j$ and additional parameters $\lbrace u \rbrace$ (e.g. Mandelstam invariants or masses)
raised to a regulated power $a_j + b_j \varepsilon$ with numeric constants $a_j$ and $b_j$. We particularly focus on Feynman parameterized loop integrals where
the $P_j$ correspond to the Symanzik polynomials $\mathcal{U}$ and $\mathcal{F}$~\cite{Heinrich:2008si}. Remember that singularities of the integrand originating from the borders
of the integration domain, i.e. where any of the $x_j$ is either zero or one\footnote{Singularities at one can be introduced by integrating out the delta function
or by the remappings of the sector decomposition procedure.},
lead to $1/\varepsilon$ poles while singularities of the integrand in the interior
of the integration domain are avoided by the Feynman prescription which is implemented via a deformation of the integration contour into the complex
plane~\cite{Soper:1999xk,Binoth:2005ff,Anastasiou:2007qb,Borowka:2012yc,Schlenk:2016cwf}.

Loop integrals usually have no singularities originating from $x_j=1$: The $\mathcal{U}$ polynomial is always positive semidefinite in the integration domain while
the kinematic dependent $\mathcal{F}$ polynomial often has a so-called Euclidean region (all Mandelstam variables negative and all masses positive),
where it is positive semidefinite as well. However, integrals without Euclidean region or a special choice of kinematics can lead to a vanishing $\mathcal{F}$ polynomial
after sector decomposition when an $x_j=1$.

Let us first consider a simple toy integral that reproduces the problem,
\begin{equation}
\lim_{\delta \rightarrow 0} \; \int_0^1 \, \mathrm{d}x \int_0^1 \, \mathrm{d}y \  \left( x - y - i \delta \right) ^ {-2 + \varepsilon},
\label{eq:xmy_integral}
\end{equation}
where the $- i \delta$ prescription is to be understood analogous to the Feynman prescription in loop integrals.
The integrand has a regulated singularity along the line $x=y$. The $- i \delta$ shifts the singularity away from the real
axis which regulates the integral in the interior of the integration domain. On the border, however, the limit $\delta \rightarrow 0$
forces the endpoints of the integral back to the real axis and thus into the singularity. Therefore, the singular endpoints $x=y=0$ and $x=y=1$
are purely regulated by the $\varepsilon$ which becomes manifest as $1/\varepsilon$ poles.

Versions one to three of \secdec{} split the integration domain as
\begin{equation}
\int_0^1 \, \mathrm{d}x_j = \int_0^{1/2} \, \mathrm{d}x_j \ + \int_{1/2}^1 \, \mathrm{d}x_j \ ,
\label{eq:secdec1_split}
\end{equation}
and then map the integration boundaries back to the unit interval by the substitutions $x_j=z_j/2$ and $x_j=1-z_j/2$, respectively.
More details about the previous split can be found in the \secdec{}-1 paper~\cite{Carter:2010hi}.

Although splitting all integration variables at $1/2$ for all integration variables seems a natural choice, that is exactly what causes unregulated singularities to remain
even after the whole sector decomposition procedure. The reason is that the singular point $x=y=1/2$ in the interior of the integration domain is mapped to an
endpoint (all $x_j$ equal to either one or zero) of the resulting integrals. However, the contour deformation vanishes at the endpoints by construction.
The singular point $x=y=1/2$ that should be avoided by a deformed contour is therefore left unregulated.
Note that the contour deformation has to vanish on the endpoints of the original integral only. Consequently, one could in principle track which new endpoints are
introduced by the split and allow for a nonzero deformation at those. One would then, however, have to match the nonzero deformation at these artificial endpoints between
all split sectors. In particular, the split sectors could then no longer be considered independent of each other.

\begin{figure}[htbp]
\hfill \\
\begin{center}
\includegraphics[width=0.5\textwidth]{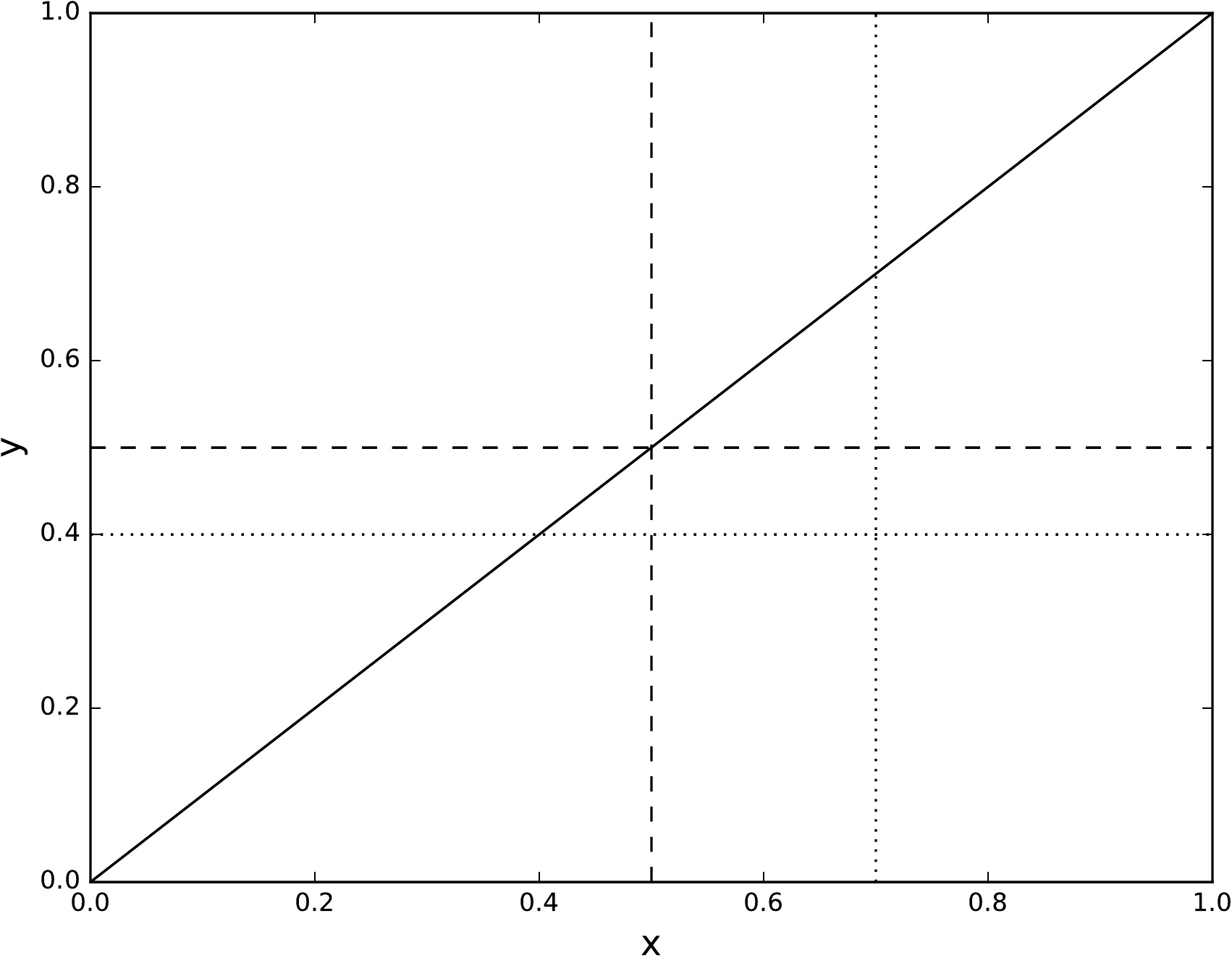}
\end{center}
\caption{Graphical representation of the ``$x - y$'' problem. The black line shows the singularity of then integrand at $x=y$.
The dashed line shows a split at $x=y=1/2$ which maps a singular point in the interior of the integration region to end points of the resulting regions.
The dotted line shows how the integrand can be split without mapping a point of the singularity to the border of the resulting integrals.}
\label{fig:xminusy}
\end{figure}

Instead, we can simply avoid mapping singular points to the border by splitting $x$ and $y$ elsewhere.
In \pysecdec{}, we generate random integers $r_j \in \left[1,19\right]$ and split the integration domain as
\begin{equation}
\int_0^1 \, \mathrm{d}x_j = \int_0^{r_j/20} \, \mathrm{d}x_j \ + \int_{r_j/20}^1 \, \mathrm{d}x_j
\label{eq:pysecdec_split}
\end{equation}
and remap the resulting integration borders back to the unit interval substituting $z_j=\frac{r_j}{20}x_j$ and $z_j = 1 - \frac{r_j}{20}x_j$, respectively.
The singularity of the integrand at $x=y$ and possible resulting integral domains from the two different splitting procedures are depicted in Figure~\ref{fig:xminusy}.

\begin{figure}[htbp]
\hfill \\
\begin{center}
\includegraphics[width=0.4\textwidth]{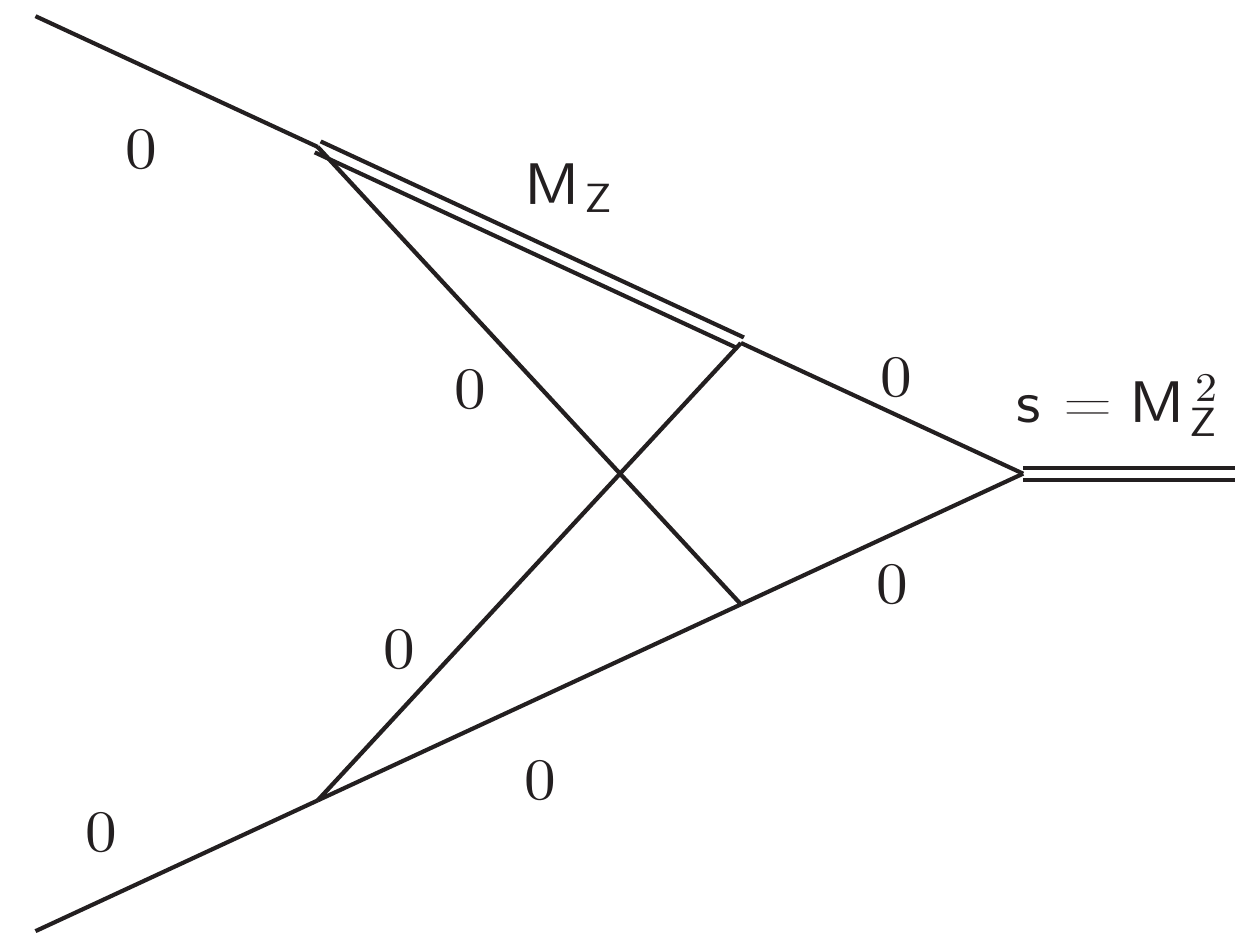}
\end{center}
\caption{A vertex diagram where splitting the integration domain as denoted in equation~\ref{eq:secdec1_split} maps a singularity in the interior
of the integration domain to an end point of the integration.
Figure taken from~\cite{Dubovyk:2016zok}.}
\label{fig:Zbb_vertex}
\end{figure}

To summarize, unregulated singularities can remain in the sector decomposed integrals if the integration domain is split such that singular points of the integrand
are mapped to endpoints of the integration. Consider for example the three-point function depicted in Figure~\ref{fig:Zbb_vertex} which is discussed in~\cite{Dubovyk:2016zok}.
Note that the squared mass of the single massive propagator is equal to the Mandelstam invariant $s$.
The $\mathcal{F}$ polynomial can be expressed as,
\begin{align}
\begin{split}
\mathcal{F}/m_Z^2 &= x_3^2 x_5 + x_3^2 x_4 + x_2 x_3 x_5 + x_2 x_3 x_4 + x_1 x_3 x_5 + x_1 x_3 x_4 \\
  &+ x_1 x_3^2 + x_1 x_2 x_3 + x_0 x_3 x_4 + x_0 x_3^2 + x_0 x_2 x_3 \\
  &- x_1 x_2 x_4 - x_0 x_1 x_5 - x_0 x_1 x_4 - x_0 x_1 x_2  - x_0 x_1 x_3.
\end{split}
\label{eq:Zbb_vertex_F}
\end{align}
where $m_Z$ is the mass of the Z-boson and the $x_j$ are Feynman parameters.

It can easily be verified that $\mathcal{F}$ vanishes for e.g. $x_0=x_1=x_2=x_4=x_5=1$ and $x_3=1/2$.
A split of $x_3$ at $1/2$ maps that singularity to the endpoints of the split integrals as described above,
which explains why this integral could not be evaluated with earlier versions of \secdec{}.

\section{Conclusion}
\label{sec:conc}

We have presented the latest updates on the development of \pysecdec{}, a toolbox
to numerically evaluate parameter integrals in the context of dimensional regularization.

We have resolved issues with integrals where the same kinematic invariant appears with opposite sign in the second Symanzik polynomial $\mathcal{F}$.

Several tests have shown that the numerical integration of sector decomposed functions can be dramatically sped up by Quasi Monte Carlo (QMC) integration and by the use
of Graphics Processing Units (GPUs). It has been shown that evaluating the master integrals numerically is a viable option for phenomenological studies.
Since the approach is general to loop integrals (and even beyond), numerically evaluating the sector decomposed master integrals may become an essential
part of automated multi-loop amplitude generators.

\section*{Acknowledgments}

I thank my fellow \secdec{} team Sophia Borowka, Gudrun Heinrich, Stephen Jones, Matthias Kerner, Johannes Schlenk,
and Tom Zirke for a successful collaboration.

\bibliographystyle{JHEP}
\bibliography{refs}

\end{document}